# Separating Use and Reuse to Improve Both

## Hrshikesh Arora[a], Marco Servetto[a], and Bruno C. d. S. Oliveira[b]

a   Victoria University of Wellington
b   The University of Hong Kong

**Abstract**   *Context:* Trait composition has inspired new research in the area of code reuse for object oriented (OO) languages. One of the main advantages of this kind of composition is that it makes possible to separate subtyping from subclassing; which is good for code-reuse, design and reasoning [15]. However, handling of state within traits is difficult, verbose or inelegant.

*Inquiry:* We identify the *this-leaking problem* as the fundamental limitation that prevents the separation of subtyping from subclassing in conventional OO languages. We explain that the concept of trait composition addresses this problem, by distinguishing code designed for use (as a type) from code designed for reuse (i.e. inherited). We are aware of at least 3 concrete independently designed research languages following this methodology: TraitRecordJ [6], Package Templates [26] and DeepFJig [16].

*Approach:* In this paper, we design $42_\mu$, a new language, where we improve use and reuse and support the **This** type and family polymorphism by distinguishing code designed for use from code designed for reuse. In this way $42_\mu$ synthesise the 3 approaches above, and improves them with *abstract state operations*: a new elegant way to handle state composition in trait based languages.

*Knowledge and Grounding:* Using case studies, we show that $42_\mu$'s model of traits with abstract state operations is more usable and compact than prior work. We formalise our work and prove that type errors cannot arise from composing well typed code.

*Importance:* This work is the logical core of the programming language 42. This shows that the ideas presented in this paper can be applicable to a full general purpose language. This form of composition is very flexible and could be used in many new languages.

## The Art, Science, and Engineering of Programming

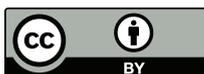

## 1 Introduction

In Java, C++, Scala and C#, subclassing implies subtyping. A Java subclass decla-ration, such as `class A extends B {}` does two things at the same time: it **inherits** code from `B`; and it creates a subtype of `B`. Therefore a subclass must *always* be a subtype of the extended class. Such design choice where subclassing implies subtyping is not universally accepted. Historically, there has been a lot of focus on separating subtyping from subclassing [15]. This separation is claimed to be good for code-reuse, design and reasoning. There are at least two distinct situations where the separation of subtyping and subclassing is helpful.

**Allowing inheritance/reuse even when subtyping is impossible:** Situations where inher-itance is desirable are prevented by the enforced subtyping relation. A well-known example are the so-called *binary methods* [13, 15]. For example, consider a class `Point` with a method `Point sum(Point o){return new Point(x+o.x,y+o.y);}`. Can we reuse the `Point` code so that `ColorPoint.sum` would take and return a `ColorPoint`? In Java/C# declaring `class ColorPoint extends Point{..}` would result in `sum` still taking a `Point` and returning a `Point`. Moreover, manually redeclaring a `ColorPoint sum(ColorPoint that)` would just induce overloading, not overriding. In this case we would like to have inheritance, but we cannot have (sound) subtyping.

**Preventing unintended subtyping:** For certain classes we would like to inherit code without creating a subtype even if, from the typing point of view, subtyping is still sound. A typical example [29] is *Sets* and *Bags*. Bag implementations can often inherit from Set implementations, and the interfaces of the two collection types are similar and type compatible. However, from the logical point-of-view a Bag is *not a subtype* of a Set.

Structural typing [15] may deal with the first situation, but not the second. Since structural subtyping accounts for the types of the methods only, a Bag would be a subtype of a Set if the two interfaces are type compatible. For dealing with the second situation, nominal subtyping is preferable: an explicit subtyping relation must be signalled by the programmer. Thus if subtyping is not desired, the programmer can simply *not* declare a subtyping relationship.

While there is no problem in subtyping without subclassing, in most nominal OO languages subclassing fundamentally implies subtyping. This is because of what we call the *this-leaking problem*, illustrated by the following (Java) code, where method `A.ma` passes `this` as a `A` to `Utils.m`. This code is correct, and there is no subtyping/subclassing.

```
1   class A{ int ma(){ return Utils.m(this); } }
2   class Utils{ static int m(A a){..} }
```

Now, lets add a class `B`:

```
1   class B extends A{ int mb(){return this.ma();} }
```

We can see an invocation of `A.ma` inside `B.mb`, where the self-reference `this` is of type `B`. The execution will eventually call `Utils.m` with an instance of `B`. However, *this can be correct only if `B` is a subtype of `A`*.

Suppose Java code-reuse (the **extends** keyword) did not introduce subtyping[1]: then an invocation of `B.mb` would result in a run-time type error. The problem is that the self-reference `this` in class `B` has type `B`. Thus, when `this` is passed as an argument to the method `Utils.m` (as a result of the invocation of `B.mb`), it will have a type that is incompatible with the expected argument of type `A`. Therefore, every OO language with the minimal features exposed in the example (using `this`, **extends** and method calls) is forced to accept that subclassing implies subtyping.

What the *this-leaking problem* shows is that adopting a more flexible nominally typed OO model where subclassing does not imply subtyping is not trivial: a more substantial change in the language design is necessary. In essence we believe that in languages like Java, classes do too many things at once. In particular they act both as units of *use* and *reuse*: classes can be *used* as types and can be instantiated; classes can also be subclassed to provide *reuse* of code. We are aware of at least 3 independently designed research languages that address the *this-leaking problem*:

- In TraitRecordJ (TR) [6, 7, 8] each construct has a single responsibility: classes instantiate objects, interfaces induce types, records express state, and traits are reuse units.
- Package Templates (PT) [2, 3, 26]: an extension of (full) Java where new packages can be "synthesized" by mixing and integrating code templates. Such "synthesized" packages can be used for code reuse without inducing subtyping.
- DeepFJig(DJ) [16, 27, 37] is a module composition language where nested classes with the same name are recursively composed.

This paper shows a simple language design, called $42_\mu$, addressing the *this-leaking problem* and decoupling subtyping from inheritance. We build on traits to distinguish code designed for *use* from code designed for *reuse*. We synthesize and simplify the best ideas from those 3 very different designs, and couple them with an elegant novel approach to state and self instantiation in traits that avoids the complexities and redundancies introduced by fields and their initialisation.

In $42_\mu$, there are two separate concepts: classes and traits [18]. Classes are meant for code *use*, and cannot be inherited/extended. Classes in $42_\mu$ are like final classes in Java, and can be used as types and as object factories. Traits are meant for code *reuse* only: multiple traits can be composed to form a class. However, traits cannot be instantiated or used as types. This allows fine-grained control of subtyping while handling examples like `Set/Bag`.

In $42_\mu$, as in many module composition languages [1], all methods can be abstract, including static ones. Moreover, module composition can be used to make an already implemented method abstract. Thus, as for dynamic dispatch, the behaviour of a method call is never set in stone. We will show how in $42_\mu$, state is induced by an implicit fixpoint operation over abstract methods, where an abstract static method can perform the role of a constructor. This allows handling examples like `Point/ColorPoint` in a natural way, without requiring code duplication.

---

[1] C++ allows "extending privately"; this is not what we mean by not introducing subtyping: in C++ it is a limitation over subtyping visibility, not over subtyping itself. Indeed, the former example would be *accepted* even if `B` were to "privately extends" `A`.

Our design brings several benefits. In particular, Family Polymorphism [20] is radically simpler to support soundly. This is already clear in the 3 lines of research above, and is even more outstanding in the clean $42_\mu$ model.

We first focus on an example-driven presentation to illustrate how to improve use and reuse. In appendix A, we then provide a compact formalization. The hard technical aspects of the semantics have been studied in previous work [2, 3, 6, 7, 8, 16, 26, 27, 37]; the design of $42_\mu$ synthesizes some of those concepts. The design ideas have been implemented in the full 42 language, which supports all the examples we show in the paper, and is available at: http://l42.is. Work on 42 is now slowly reaching maturity after about 5 years of intense research and development. The current implementation is now robust enough to create realistic medium sized programs running on the JVM, and the standard library consists of over 10000 lines of 42 code.

In summary, our contributions are:

- We identify the this-leaking problem, that makes separating inheritance and subtyping difficult.
- We synthesize the key ideas of previous designs that solve the this-leaking problem into a novel and minimalistic language design. This language is the core logic of the language 42, and all the examples in this paper can be encoded as valid 42 programs. This design improves both code use and code reuse.
- We propose a clean and elegant approach to the handling of state in a trait based language.
- We illustrate how $42_\mu$, extended with nested classes, enables a powerful (but at the same time simple) form of family polymorphism.
- We show the simplicity of our approach by providing a compact 1 page formalization (in Appendix A).
- We perform 3 case studies, comparing our work with other approaches, and we collect clear data showing that we can express the same examples in a cleaner and more modular manner.

## 2 The Design of $42_\mu$: Separating Use and Reuse

### 2.1 Classes in $42_\mu$: a mechanism for code use

Consider the example of section 1 rewritten in $42_\mu$, introducing classes `Utils` and `A`:

```
1  A={ method int ma(){ return Utils.m(this); } }
2  Utils={ static method int m(A a){ return ..; } }
```

Classes in $42_\mu$ use a different declaration style compared to Java: there is no `class` keyword, and an equals sign separates the class name (which must always start with an uppercase letter) and the class implementation, which is used to specify the definitions of the class. In our example, in the class declaration for `A`, the name of the class is `A` and the code literal associated with the class is '{ **method int** ma(){**return** Utils.m(**this**);}}' and it contains the method ma(). In the $42_\mu$ code above, there is no way to add a class B reusing the code of A: class A (uppercase) is designed for code *use* and not *reuse*. Indeed, a noticeable difference with Java is the absence of the **extends** keyword. $42_\mu$ classes are roughly equivalent to final classes in Java. This means that there is actually no subclassing. Thus, unlike the Java code, introducing a subclass B is not possible. This may seem like a severe restriction, but $42_\mu$ has a different mechanism for *code-reuse* that is more appropriate when *code-reuse* is intended.

## 2.2 Traits in $42_\mu$: a mechanism for code reuse

Traits in $42_\mu$ cannot be instantiated and do not introduce new types. However they provide code reuse. Trait declarations look very much like class declarations, but trait names start with a lowercase letter (even syntactically they can not be used as types).

```
1  Utils={ static method int m(A a){return ...} }
2  ta={ method int ma(){return Utils.m(this);}} //type error
3  A=Use ta
```

Here ta is a trait intended to replace the original class A so that the code of the method ma can be reused. Then the class A is created by reusing the code from the trait ta, introduced by the keyword **Use**. Note that **Use** expressions cannot contain class names: only trait names are allowed. *Referring to a trait is the only way to induce code reuse*.

The crucial point is the call Utils.m(**this**) inside trait ta: the corresponding call in the Java code is correct since Java guarantees that such occurrence of **this** will be a subtype of A everywhere it is reused. In $42_\mu$ the type of **this** in ta has no relationship to the type A; thus the code Utils.m(**this**) is illtyped.

The following second attempt would not work either:

```
1  Utils={ static method int m(ta a){return ...}} //syntax error
2  ta={ method int ma(){return Utils.m(this);}}
3  A=Use ta
```

ta is not a type in the first place, since it is a (lowercase) trait name. Indeed, trait names can only be used in **Use** expressions, and thus they can not appear in method bodies or type annotations. In this way, the code of a trait can stay agnostic of its name. This is one of the key design decisions in $42_\mu$: traits can be reused in multiple places, and their code can be seen under multiple types. In $42_\mu$, *interfaces are the only way to obtain subtyping*. As shown in the code below, interfaces are special kinds of code literals, where all the methods are abstract. Thus, to model the original Java example, we need an interface capturing the commonalities between A and B:

```
1  IA={interface method int ma()} //interface with abstract method
2  Utils={static method int m(IA a){return ...} }
3  ta={implements IA //This line is the core of the solution
4      method int ma(){return Utils.m(this);}}
5  A=Use ta
```

This code works: Utils relies on interface IA and the trait ta implements IA. Any class reusing ta will contain the code of ta, including the **implements** IA subtyping declaration; thus any class reusing ta will be a subtype of IA. Therefore, while typechecking Utils.m(**this**) we can assume **this**<:IA. It is also possible to add a class B as follows:

```
1    B=Use ta, { method int mb(){return this.ma();} }
```

This also works. B reuses the code of ta, but has no knowledge of A. Since B reuses ta, and ta implements IA, B also implements IA.

Later, in appendix A we will provide the type system. Here notice that the former declaration of B is correct even if no method called ma is explicitly declared. DJ and TR would instead require explicitly declaring an abstract ma method:

```
1    B=Use ta, { method int ma() //not required by us
2        method int mb(){return this.ma();} }
```

In 42$_\mu$, methods are directly accessible from ta, exactly as in the Java equivalent

```
1    class B extends A{ int mb(){return this.ma();} }
```

where method ma is imported from A. This concept is natural for a Java programmer, but was not supported in previous work [8, 16]. Those works require all dependencies in code literals to be explicitly declared, so that the code literal is self-contained; in this way it can be typed in isolation before flattening. However, this results in many redundant abstract method declarations.

**Semantics of Use:**  The semantics of traits is defined with flattening, which is simple to formalize and understand. However, if implemented naively, flattening may cause a lot of bytecode duplication. The 'delegation semantics' [28], is a proposed alternative semantic model for traits that is observationally equivalent to flattening but does not require bytecode duplication. The formalism presented here will rely on simple flattening, but we expect the techniques of [28] would be useful to produce an efficient implementation in term of bytecode space. With the flattening semantics A and B are equivalent to the inlined code of all used traits.

```
1    A=Use ta
2    B=Use ta, { method int mb(){return this.ma();} }
3    //equivalent to
4    A={implements IA method int ma(){return Utils.m(this);}}
5    B={implements IA
6        method int ma(){return Utils.m(this);}
7        method int mb(){return this.ma();} }
```

*In the resulting code, there is no mention of the trait ta.* Information about code-reuse/inheritance is a private implementation detail of A and B; while subtyping is part of the class interface.This position has been defended by Bracha [10]: the choice of inheriting behaviour should be in the hands of the programmer; if a method implementation is not appropriate, such method can be overridden. If too many methods do not provide an appropriate behaviour, inheriting code from another location or implementing the behaviour from scratch may also be considered. This should not impact the interface exposed to the user, otherwise the programmer may be unable to change their implementation decisions at a later time. In summary, to leak **this** in 42$_\mu$, either code reuse is disallowed, or an appropriate interface (IA in this case) must be implemented. We believe the code with IA better transmits programmer intention. Some readers may instead see requiring IA as a cost of our approach. Even from this point of view, such cost is counter balanced by the very natural and simple support for code reuse, '**This**' type and (in the extensions with nested classes seen later) family polymorphism. The syntactic cost of introducing new names can be reduced with some syntactic sugar.

## 3 Improving Use

To illustrate how $42_\mu$ improves the use of classes, we model a simplified version of Set and Bag collections first in Java, and then in $42_\mu$. The benefit of $42_\mu$ is that we get reuse without introducing subtyping between Bags and Sets. As shown below, this improves the use of Bags by eliminating logical errors arising from incorrect subtyping relations that are allowed in the Java solution.

### 3.1 Sets and Bags in Java: the need for code reuse without subtyping

An iconic example on why connecting inheritance/code reuse and subtpying is problematic is provided by LaLonde [29]. A reasonable implementation for a `Set` is easy to extend into a `Bag` by keeping track of how many times an element occurs. We just add some state and override a few methods. For example in Java one could have:

```
1  class Set {..//usual hashmap implementation
2    private Elem[] hashMap;
3    void put(Elem e){..}
4    boolean isIn(Elem e){..}}
5  class Bag extends Set{ ..//for each element in the hash map,
6    private int[] countMap;// keep track of how many occurrences are in the bag
7    @Override void put(Elem e){..}
8    int howManyTimes(Elem e){..}}
```

Coding `Bag` in this way avoids a lot of code duplication, but we induced unintended subtyping! Since subclassing implies subtyping, our code breaks the Liskov substitution principle (LSP) [30]: not all bags are sets![2] Indeed, the following is allowed:

```
1  Set mySet=new Bag(); //OK for the type system but not for LSP
```

This encumbers the programmer: to avoid conceptual errors that are not captured by the type system, they have to *use* `Bag` very carefully.

**A (broken) attempt to fix the Problem in Java:** One could *retroactively* fix this problem by introducing `AbstractSetOrBag` and making both `Bag` and `Set` inherit from it:

```
1  abstract class AbstractSetOrBag {/*old set code goes here*/}
2  class Set extends AbstractSetOrBag {} //empty body
3  class Bag extends AbstractSetOrBag {/*old bag code goes here*/}
4  ...
5  //AbstractSetOrBag type not designed to be used.
6  AbstractSetOrBag unexpected=new Bag();
```

---

[2] The LSP is often broken in real programs because of the need of inheritance: the LSP allows only refinement not extension. Traits provide extension without breaking the LSP.

This looks unnatural, since `Set` would extend `AbstractSetOrBag` without adding anything, and we would be surprised to find a use of the type `AbstractSetOrBag`. Worst, if we are to constantly apply this mentally, we would introduce a very high number of abstract classes that are not supposed to be used as types. Those classes would clutter the public interface of our classes and the project as a whole. A *use*able API should provide only the information relevant to the client. In our example, the information `Set<:AbstractSetOrBag` would be present in the public interface of the class `Set`, but such information is not needed to use the class properly! Moreover, the original problem is not really solved, but only moved further away. For example, one day we may need bags that can only store up to 5 copies of the same element. We are now at the starting point again:

- either we insert **class** `Bag5` **extends** `Bag` and we break the LSP;
- or we duplicate the code of the `Bag` implementation with minimal adjustments in **class** `Bag5` **extends** `AbstractSetOrBag`;
- or we introduce an **abstract class** `BagN` **extends** `AbstractSetOrBag` and **class** `Bag5` **extends** `BagN` and we modify `Bag` so that **class** `Bag` **extends** `BagN`. Note that this last solution is changing the public interface of the formerly released `Bag` class, and this may even break backwards-compatibility (if a client program was using reflection, for example).

### 3.2 Sets and Bags in 42$_\mu$

Instead, in 42$_\mu$, if we were to originally declare

```
1  Set={/*set implementation*/}
```

Then our code would be impossible to reuse in the first place for any user of our library. We consider this an advantage, since unintended code reuse runs into under-documented behaviour nearly all the time![3] If the designer of the `Set` class wishes to make it reusable, they can do it explicitly by providing a set trait:

```
1  set={/*set implementation*/}
2  Set=Use set
```

Since `set` can never be used as a type, there is no reason to give it a fancy-future-aware name like `AbstractSetOrBag`. There are two different ways to add the concept of bags:

```
1  set={/*set implementation*/} //version 1
2  Set=Use set
3  Bag= Use set, {/*bag implementation*/}
4
5  set={/*set implementation*/} //version 2
6  Set=Use set
7  bag=Use set, {/*bag implementation*/}
8  Bag=Use bag
```

---

[3] See "Design and document for inheritance or else prohibit it" [9]: the self use of public methods is rarely documented, thus is hard to understand the effects of overriding a library method.

Notice how, thanks to flattening, the resulting code for `Bag` is identical in both versions and, as shown in Section 2, there is no trace of trait `bag` at run time. Thus if we are the developers of bags, we can temporarily go for the first version. Then, when for example we need to add `Bag5` as discussed before, we can introduce the `bag` trait without adding new undesired complexity for our old clients.

## 4 Improving Reuse

To illustrate how $42_\mu$ improves reuse, we show a novel approach to smoothly integrating state and traits: a challenging problem that has limited the flexibility of traits and reuse in the past. The idea of flattening is elegant and successful in module composition languages [1] and several trait models [5, 8, 18, 27]. Flattening is elegant in these two settings since traits (or modules) only have one kind of member: methods (or functions). In this way flattening is defined as simply collecting all members from all used traits (or composed modules), where methods with same name and type signature are summed into a single one. At most one of those summed methods can have a body, which will be propagated into the result. However the research community is struggling to make it work with object state (constructors and fields) while achieving the following goals:

- managing fields in a way that borrows the elegance of summing methods;
- actually initializing objects, leaving no null fields;
- making it easy to add new fields;
- allowing self instantiation: a trait method can instantiate the class using it.

An in-depth discussion on how such goals are difficult to achieve and how they have been challenged in the existing literature is available in Section 7.3.

### 4.1 State of the art

We first present the state of the art solution: traits have only methods but classes also have fields and constructors. The idea is that the trait code just uses getter/setters/factories, while leaving classes to finally define the fields/constructors. That is, in this state of the art solution, classes have a richer syntax than traits, allowing declaration for fields and constructors.

**Points:** Consider two traits dealing with *point* objects with coordinates `x` and `y`.

```
1  //idealized state of the art trait language, not 42
2  pointSum= { method int x() method int y()//getters
3    static method This of(int x,int y)//factory method
4    method This sum(This that){//sum code
5      return This.of(this.x()+that.x(),this.y()+that.y());//self instantiation
6    }}
7  pointMul= { method int x() method int y()//repeating getters
8    static method This of(int x,int y)//repeating factory
9    method This mul(This that){//multiplication code
10     return This.of(this.x()*that.x(),this.y()*that.y()); }}
```

The first trait provides a *binary method* that adds the point object to another point to return a new point. The second trait provides multiplication. In this code all the operations dealing with state are represented as *abstract methods*. Notice the abstract **static method This** of(..) which acts as a factory/constructor for points. As for instance methods, static methods are late bound: flattening can provide an implementation for them. Thus, in $42\mu$ they can be abstract, and abstract static methods are similar to the concept of member functions in the module composition setting [1]. Following the traditional model of traits and classes common in literature [18], we can compose the two traits, by *adding glue-code* to implement methods x, y and of. This approach is verbose but very powerful, as illustrated by ClassLess Java [40].

```
1  //idealized state of the art language, not 42
2  class PointAlgebra=Use pointSum,pointMul, {//not 42 code
3      int x int y//unsatisfactory state of the art solution
4      constructor PointAlgebra(int x, int y){ this.x=x this.y=y }
5      method int x(){return x;}//repetitive code
6      method int y(){return y;}// in traits terminology, this is all "glue code"
7      static method This of(int x, int y){return new PointAlgebra(x,y);} }
```

With a slightly different syntax, this approach is available in both Scala and Rust, and they both require glue code. It has some advantages, but also disadvantages:

**Advantages:** This approach is associative and commutative, even self instantiation can be allowed if the trait requires a static method returning **This**. The class will then implement the methods returning **This** by forwarding a call to the constructor.

**Disadvantages:** The class needs to handle all the state, even state conceptually private to a trait. Moreover, writing such obvious code to close the state/fixpoint in the class with the constructors and fields and getter/setters and factories is tedious and error prone; such code could be automatically generated [40].

### 4.2 Our proposed approach to State: Coherent Classes

In $42\mu$ there is no need to write down constructors and fields. In fact, in $42\mu$ there is not even syntax for those constructs! The intuition is that a class where *all* abstract methods can be seen as field getters, setters, or factories, is a *coherent* class. In most other languages, a class is abstract if it has abstract methods. Instead, we call a class abstract only when the set of abstract methods are not coherent. That is, the abstract methods cannot be automatically recognised as factory, getters or setters. Methods recognised as factory, getters and setters are called *abstract state operations*.

A definition of coherent classes is given next, and is formally modelled in appendix A:

- A class with no abstract methods is coherent (just like Java Math, for example). Such classes have no instances and are only useful for calling static methods.
- A class with a single abstract **static** method returning **This** and with parameters $T_1\ x_1, \ldots, T_n\ x_n$ is coherent if all the other abstract methods can be seen as *abstract state operations* over one of $x_1, \ldots, x_n$. That is:
  - A method $T_i\ x_i()$ is interpreted as an abstract state method: a *getter* for $x_i$.
  - A method **void** $x_i(T_i\ \text{that})$ is a *setter* for $x_i$.

Note how the single, abstract static method acts as a *factory method*. The signature of the factory method plays an important role, since abstract state operations are identified by using the names of the factory method arguments. The idea of creating objects in a single atomic step by providing a value for all their fields is well explored (such as with primary constructors in Scala) and does not limit the freedom of programmers to specify personalised initialisation strategies. A static method can freely compute concrete field values before creating objects. Appendix B.4 discusses usability implications of this pattern.

While getters and setters are fundamental operations, it is possible to support more operations. For example:

- **method This** withX(**int** that) may create a new instance that is like **this** except that field x now has value that. Those kinds of methods performs functional field updates and are called *withers*.
- **method This** clone() may do a shallow clone of the object.

The concept of 'abstract state operations' is novel, and we think it is a promising area for further research. ClassLess Java [40] explores a particular set of such abstract state operations, but we suspect there are more unexplored possible options that could be even more beneficial.

**Points in $42_\mu$:**   In $42_\mu$ and with our approach to handle the state, pointSum and pointMul can indeed be directly composed. This works because the resulting class is coherent.

```
1  PointAlgebra= Use pointSum,pointMul //no glue code needed
```

**Improved solution:**   So far the current solution still repeats the abstract methods x, y and of. Moreover, in addition to sum and mul we may want many operations over points. It is possible to improve reuse and not repeat such declaration by abstracting the common declaration into a trait p:

```
1   p= { method int x() method int y()
2      static method This of(int x,int y)
3      }
4   pointSum= Use p, {
5      method This sum(This that){
6         return This.of(this.x()+that.x(),this.y()+that.y());
7      }}
8   pointMul= Use p, {
9      method This mul(This that){
10        return This.of(this.x()*that.x(),this.y()*that.y());
11     }}
12  pointDiv= ...
13  PointAlgebra= Use pointSum,pointMul,pointDiv,...
```

Now the code is fully modularized, that is: each trait defines exactly one method and contains its abstract dependencies. In this way it can be modularly composed with any code requiring such a method.

**Case Study 1:**   In order to evaluate our approach we performed a case study: we consider 4 different operations Sum, Subtraction, Multiplication and Division. These operations can be combined in 16 different ways. We wrote this example in four different styles: (a) Java7 *(115 lines)*, (b) Classless Java *(82 lines)*, (c) Scala *(81 lines)* and (d) $42\mu$ *(32 lines)*.[4] We chose Classless Java [40] since it is a novel approach allowing Java8 default interface methods to encode traits in Java. We then chose Java7, that lacks the features needed to encode traits, to show the impact of this feature. Finally, the comparison with Scala is interesting since it has good support for traits, and using abstract types, it is possible to support the '**This**' type. Rust is similar to Scala in this regard; we believe we would get similar results by comparing against either Scala or Rust.

| Language | Lines of code | members | classes/traits |
|---|---|---|---|
| Java7 | $115 = 6 + 5 * 4 + 7 * 6 + 9 * 4 + 11$ | 50 | 16 |
| Classless Java | $82 = 3 + 3 * 4 + 5 * 6 + 7 * 4 + 9$ | 34 | 16 |
| Scala | $81 = 5 + 3 * 4 + 4 * 16$ | 40 | $21 = 16 + 4 + 1$ |
| $42\mu$ | $32 = 4 + 3 * 4 + 1 * 16$ | 7 | $21 = 16 + 4 + 1$ |

We observed that in Java7 we had to duplicate[5] 28 method bodies across the 16 classes. Of these, 11 method bodies were duplicated because Java does not support multiple inheritance and the remaining 17 bodies had to be duplicated to ensure that the right type is returned by the method. Those could be avoided if Java supported the '**This**' type. On the other hand, the solution in $42\mu$ was much more compact since we could efficiently reuse traits (this is why the number of top-level concepts in $42\mu$ was larger i.e. 21 due to the presence of traits in this solution). In detail, Java required 6 lines for the initial `Point` class, 5 lines for each of the 4 arithmetic operations, 7 lines for each of the 6 combinations of two different operations, 9 lines for each of the 4 combinations of three different operations and finally 11 lines for the class with all four operations. The solution in Classless Java was slightly smaller than Java7, but was still longer than the $42\mu$ solution: it still had to redefine the sum, sub and other operations in each of the classes. Here the limited support for the '**This**' type is to blame, thus Classless Java also has 28 duplicated method bodies.

Finally, we compare it with a Scala solution. There is no need for duplicate method bodies in Scala. However, for '**This**' instantiation we need to define abstract methods, that will be implemented in the concrete classes. The Scala solution has the same exact advantages of our proposed solution, and the declaration of the trait is about the same size: 5 (point state) $+3 * 4$ (point operations). However the glue code (the code needed to compose the traits into usable classes) is quite costly: 4 lines for each of the 16 cases. In $42\mu$ a single line for each case is sufficient.

This example is the best-case scenario for $42\mu$: where a maximum level of reuse is required since we considered the case where all the 16 permutations needed to be materialized in the code. In all our case studies, to make a meaningful comparison, we formatted all code in a readable and consistent manner; on the other hand for space limitations, the code snippets presented in the article are formatted for compactness.

---

[4] Since we want to focus on the actual code, while counting line numbers we *omit* empty lines and lines containing only open/closed parenthesis/braces.

[5] A duplicate body is repetition of identical code (may have different types in its scope/environment). The first occurrence is not counted.

### 4.3 State Extensibility

Programmers may want to extend points with more state. For example they may want to add colors to the points. A first attempt at doing this would be:

```
1 colored= { method Color color() }
2 CPoint= Use pointSum,colored //Fails: class not coherent
```

This first attempt does not work: the abstract color method is not a getter for any of the parameters of **static method This** of(**int** x,**int** y). A solution is to provide a richer factory:

```
1 CPoint= Use pointSum,colored,{
2    static method This of(int x,int y){return This.of(x,y,Color.of(/*red*/));}
3    static method This of(int x,int y,Color color) }
```

where we assume support for overloading based on different numbers of parameters. This is a reasonable solution, however the method CPoint.sum resets the color to red: we call the of(**int**, **int**) method, that now delegates to of(**int**, **int**, Color) by passing red as the default field value. What should be the behaviour in this case? If our abstract state supports withers, we can use this.withX(newX).withY(newY), instead of writing **This**.of(...), in order to preserve the color from this. This solution is better but still not satisfactory since the color from that is ignored.

**A better design:** We can design trait p for reuse and extensibility by adding an abstract merge(**This**) method as an extensibility hook; colored can now define color merging. Using withers we can merge colors, or any other kind of state following this pattern.

```
1 p= { method int x() method int y() //getters
2    method This withX(int that) method This withY(int that)//withers
3    static method This of(int x,int y)
4    method This merge(This that) //new method merge!
5    }
6 pointSum= Use p, {
7    method This sum(This that){
8       return this.merge(that).withX(this.x()+that.x()).withY(this.y()+that.y());
9    }}
10 colored= {method Color color()
11    method This withColor(Color that)
12    method This merge(This that){ //how to merge color handled here
13       return this.withColor(this.color().mix(that.color()));
14    }}
15 CPoint= /*as before*/
```

**Independent Extensibility:** Of course, quite frequently there can be multiple independent extensions [41] that need to be composed. Lets suppose that we could have a notion of flavored as well. In order to compose colored with flavored we would need to compose their respective merge operations. To this aim **Use** is not sufficient. To combine the implementation of two different implementation of methods, we introduce an operator called **super**, that makes a method abstract and moves the implementation to another name. This is very useful to implement super calls and to compose conflicting implementations. Consider the simple flavored trait:

```
1  flavored= {
2    method Flavor flavor() //very similar to colored
3    method This withFlavor(Flavor that)
4    method This merge(This that){ //merging flavors handled here
5      return this.withFlavor(that.flavor());}}//inherits "that" flavor
```

To merge colored and flavored we use **super** to introduce method selectors _1merge and _2merge to refer to the version of merge as defined in the first/second element of **Use**.

```
1  FCPoint= Use
2    colored[super merge as _1merge], //this leaves merge as an abstract method, and
3    flavored[super merge as _2merge],//copies the bodies into _1merge and _2merge
4    pointSum,{
5      static method This of(int x,int y){
6        return This.of(x,y,Color.of(/*red*/),Flavor.none());}
7      static method This of(int x, int y,Color color,Flavor flavor)
8      method This merge(This that){//merge conflict is solved
9        return this._1merge(that)._2merge(that);} }//by calling the two versions
```

We are relying on the fact that the code literal does not need to be complete, thus we can just call _1merge and _2merge without declaring their abstract signature explicitly.

In this last example, when we tried to obtain state extensibility, we refactored the code to introduce the merge(**This**) method. This suggests that we had to anticipate the need for state extensibility in order to design our original code. As illustrated by the following example, we can instead rely on the **super** operator to inject the merge(**This**) method when needed.

```
1  p=/*as originally designed: no merge*/
2  pointSum=/*as originally designed: no merge*/
3  merge={method This merge(This that)}
4  pointSumMerge=Use merge, pointSum[super sum as _1sum], {
5    method This sum(This that){return this.merge(that)._1sum(that);}}
6  colored=/*as before, with merge implementation*/
7  CPoint= /*as before, but using pointSumMerge*/
```

**Case Study 2:**    To understand how easy it is to extend the state in this way we compare the former code with an equivalent version in Java. For this example, in Java we encode Point with the fields but no operations, PointSum reuses Point adding a functional sum operation, CPoint reuses PointSum with a Color field and FCPoint reuses CPoint with a Flavour field. This second case study represents a *worst case scenario* for $42_\mu$ against Java because we model just a single chain of reuse, easily supported in plain Java by single inheritance. Like the previous experiment, we still found that the Java solution was longer (47 lines) than that in $42_\mu$ (33 lines). This is caused by the absence of support for the '**This**' type, where the withers in each of the CPoint/FCpoint classes had to be repeated to make sure that the returned type will be correct (the number of members in Java were 27 while 24 (3 less) in $42_\mu$).

Complex patterns in Java[6] allow supporting the '**This**' type and '**This**' type instantiation but they require a lot of set-up code. We experimented with those patterns, but it soon became very clear that the resulting code of this approach would have been even larger; albeit without duplicated code. Note how the Java code is less modular than the $42_\mu$ code, since Colored and Flavored do not exist as individual concepts.

We also compare with a solution in Scala, offering the same level of reuse and code modularity of the $42_\mu$ solution, but again it is more verbose and requires more members (31): an indication that it may be logically heavier too. We define the main tPoint trait (8 lines), the tPointSum operation (3), the two tColored and tFlavored traits ($6 * 2$) and the CPoint and CFPoint classes ($12 + 18$). The major benefit of $42_\mu$ is the reduction of the amount of glue-code needed to generate CPoint and CFPoint ($4 + 9$). The results for the second experiment are presented below.

| Language | Lines of code | | members | classes or traits |
|---|---|---|---|---|
| Java | $47 = 10 + 9$ | $+$ | $13 + 15$ | 27 | 6 |
| Scala | $53 = 8 + 3 + 6 * 2 + 12 + 18$ | | 31 | 6 |
| $42_\mu$ | $33 = 7 + 3 + 5 * 2 + 4 + 9$ | | 24 | 6 |

## 5  Family Polymorphism by Disconnecting Use and Reuse

A nested class is just another kind of member in a code literal. In Java and Scala if a subclass declares a nested class with the same name of a nested class of a superclass, the parent declaration is simply hidden. The main idea of family polymorphism (FP) [11, 16, 21, 24, 25, 32, 36] is to instead consider such definition a form of *overriding*, called *further extension*. That is, the following Java code is ill typed:

```
1 abstract class A{static class B{..} abstract B m();}
2 class AA extends A{static class B{..} B m(){..}}//Error: Invalid overriding
3 //method AA.m() return type is AA.B, that is unrelated to A.B
```

In the FP approach, class AA.B would further extend A.B, thus the overriding of method A.m would be accepted. We extend $42_\mu$ with nested classes, so that by composing code with **Use**, nested classes with the same name are recursively composed. The corresponding code in $42_\mu$ would work, and behave like further extension in FP.

```
1 a={B={..} method B m()}
2 AA= Use a, {B={..} method B m(){..}}
```

For simplicity, we discuss nested classes but not nested traits: and all traits and code composition expressions are still at top level. In this way all dependencies are about top level names, allowing the type system to consider the class table as a simple map from (nested) type names (such as A and A.B.C) to their definition.

There are a lots of different forms of rename in literature [1, 10, 16]. Here we introduce a simple variant to rename nested types to other nested types. For example:

---

[6] Combining the ones used in those works [34, 38], with abstract methods to allow self instantiation as in [41].

```
1  t={ method B m()    B={ method B mb()} }
2  D= t[rename B into C]
3  //the code above flattens to the following:
4  t={ method B m()    B={ method B mb()} }
5  D={ method C m()    C={ method C mb()} }
```

The rename only influences its argument. Since traits do not induce nominal types, we can consistently change their internally used names without breaking any code. The full L42 offers many other kinds of renames, but we do not need them to show our next example.

**Application to the expression problem. Case Study 3:**

The above extensions lets us challenge the expression problem [39], with the requirements exposed in [41]. In the expression problem we have data-variants and operations and we can *extend our solution in both dimensions*, by adding new data-variants and operations. We aim to *combine independently developed extensions* so that they can be used jointly. To be modular, extensions will preserve *type safety* and allow *separate compilation* (no re-type-checking), while avoiding *duplication of source code*.

Following closely the example of Zenger and Odersky [41], we consider a language where the expressions Exp can be Num (for number literal), Plus (for binary plus operator) and Neg (for unary minus). We then proceed to define operations show to convert them into strings, eval to compute their numeric values and double to double their containing Nums. We thus have 3 classes, 1 interface, the definition of the state, and 3 operations. We model this as a table of features, as in [16]: a (3 classes + 1 interface)*(1 state + 3 operations) table composed by 16 traits. The features are atomic: they exactly declare the state of a class or define a single operation for a single class. $42_\mu$ avoids the large amount of abstract declarations that clutters the solution in [16]. Intuitively, we would like our traits to look like the following:

```
1  evalPlus= Use plus, {//eval operation for Plus data-variant
2    Exp= {interface
3      method int eval()}
4    Plus= {implements Exp
5      method int eval(){return this.left().eval()+this.right().eval();}}}
```

evalPlus uses the trait plus to import the state (the left() and right() methods) and defines the eval() method from interface Exp. But, if we were to declare those explicitly, we would repeat Exp, the abstract declaration of eval() and 'implements Exp' for all data-variants. To avoid this duplication, we write the trait eval with a placeholder T nested class, that can then be renamed into the corresponding data-variant. Thus, our source code is as follows; First we declare the 4 traits to represent the state:

```
1  exp= {    Exp= {interface}    T= {implements Exp}}
```

```
1  num= Use exp[rename T into Num],{//T is renamed to Num and summed with
2    Num= {method int value() static method Num of(int value)}} // this Num
```

```
1  plus= Use exp[rename T into Plus], {
2    Plus= {method Exp left() method Exp right()
3      static method Plus of(Exp left,Exp right)}}
```

```
1  neg= Use exp[rename T into Neg],{
2    Neg= {method Exp term() static method Neg of(Exp term)}}
```

Here we define a trait for each data-variant. Each trait will contain its version of `Exp` and a specific kind of expression, with its state. Next, we define the operation `eval` for all the data-variants. The former solutions in [16] required repeating the state declaration of the data-variant in each operation, while we can just import it.

```
1  eval= {Exp= {interface     method int eval()}    T= {implements Exp}}
```

```
1  evalNum= Use num, eval[rename T into Num],{ //just the implementation
2    Num= { method int eval(){return this.value();} }}//of the specific method
```

```
1  evalPlus= Use plus, eval[rename T into Plus], {
2    Plus= { method int eval(){ return this.left().eval()+this.right().eval(); }}
```

```
1  evalNeg= Use neg, eval[rename T into Neg], {Neg= { method int eval(){..}}}
```

The `show` operation can be trivially defined following exactly the same pattern (omitted here for space reasons). The operation `double` is a challenge for some proposed solutions to the expression problem, as explained by Zhang and Oliveira [42]. The `double` operation is called a *transformation*: an operation from `Exp` to `Exp`. Thanks to $42_\mu$'s separation between use and reuse, together with support for self-instantiation of nested classes `double` does not need any special attention and can be written just like `eval` and `show`.

```
1  double= {   Exp= {interface     method Exp double()}    T= {implements Exp} }
```

```
1  doubleNum= Use num, double[rename T in Num],{
2    Num= { method Exp double(){return Num.of(this.value()*2);} }}
```

```
1  doublePlus= Use plus,double[rename T in Plus],{
2    Plus= { method Exp double(){
3      return Plus.of(this.left().double(),this.right().double());} }}
```

```
1  doubleNeg=....
```

Here we define a trait for each data-variant implementing the operation `double()`. Again, each trait will contain its version of `Exp` with `double()` and a specific kind of expression, with the implementation for `double()` for that specific kind.

Our third case study compares with the results presented in Scala [41]. The proposed solution is not fully modularized as a table, so in order to make a more close comparison, we provide an alternative version where we isolate all the units of behaviour as is done in $42_\mu$.

|  | lines | methods |
|---|---|---|
| Original Scala | 52 | $15 = 12 + 3$ |
| Scala isolated units | 78 | $15 = 12 + 3$ |
| Scala glue-code | 27 | 3 |
| 42 traits | 48 | $19 = 4 \times 3 + 7$ |
| 42 classes | 3 | 0 |

Scala uses $12 = 4 \times 3$ methods plus 3 extra factory methods (for `double`). We use $12 = 4 \times 3$ methods plus our abstract state: 4 getters and 3 factories. As we can see, encoding atomic units in Scala is more verbose, but more importantly, in $42_\mu$ we can just define a class supporting any subset of operations and data-variants by listing the desired traits: for example, a solution for `Num` and `Plus` (but not `Neg`) with `eval` and `double` would look like this: `Example= Use evalNum,evalPlus,doubleNum,doublePlus`. The composition of all our traits would just requiring listing all of the relevant behaviour; reasonably formatted, it could take up to 3 lines. On the other hand, the presented solution in Scala requires 27 lines of glue code to put the traits together. This means that a full Scala solution *requiring a single instantiation with all the traits* would be $78 + 27 = 105$ lines. If we were to require more instantiations with a different subset of traits, the glue code would dominate the line count, and the Scala solution would end up being up to 9 times heavier than the $42_\mu$ one (if all 64 permutations were required).

The line count for $42_\mu$ is very predictable: after defining exp (3) and the state traits $(4 + 6 + 5)$ for each of the three operations (eval,show,double) we just needed 4 lines to declare the operation in the interface, and 2 lines for each of the 3 data-variants.

Following [41], after double we present an implementation of equals. Their solution involved double dispatch to avoid casting. To show understandable code, we show a simpler solution with a guarded cast (sometime called a typecast).[7] The idea is that since every data-variant contains the same "cast" logic, we can modularize it into an equals trait; equals in [41] is complex and and requires glue code.

```
1  equals= {
2    Exp= {interface    method Bool equals(Exp that)}
3    T= {implements Exp
4      method Bool exactEquals(T that)
5      method equals(that){
6        if(T instanceof This){return this.exactEquals(that);}else{ return false;}}}
```

```
1  equalsNum= Use num, equals[rename T into Num],{
2    Num= {method Bool exactEquals(Num that){
3      return this.value().equals(that.value());} }}
```

```
1  equalsPlus= Use plus, equals[rename T into Plus],{
2    Plus= {method Bool exactEquals(Plus that){
3      return this.left().equals(that.left()) && this.right().equals(that.right());
4    } }}
```

```
1  equalsNeg= Use neg, equals[rename T into Neg],{
2    Neg= {method Bool exactEquals(Neg that){
3      return this.term().equals(that.term());} }}
```

|  | lines | methods |
|---|---|---|
| Original Scala equals | 40 | 10 |
| Isolated Scala equals | 31 | 10 |
| Scala equals instance | 29 | 3 |
| 42 trait eq d-dispatch | 21 | 6 |
| 42 class dd instance | 22 | 11 |
| 42 traits eq Cast | 13 | 6 |
| 42 class cast instance | 3 | 0 |

The Scala code here can be made fully isolated with little extra syntactic cost. The original Scala eq is 40 lines and contains a part of the glue code mixed inside. The isolated version is 31 lines and to merge all the operations together in Scala, it takes 29 lines of glue code. Note that this is mostly the same glue code from before (27 lines), that needs to be manually adapted.

In $42_\mu$ we are more compact than Scala both when using the double dispatch (21+22

---

[7] The interested reader can find a $42_\mu$ implementation of equals with double dispatch in the appendix.

vs. $31+29$) or the guarded cast ($13+3$ vs. $31+29$). To instantiate the double dispatch version in $42_\mu$ we need 22 lines of glue code. We could remove such glue code using features from the full 42 language, but here we stick to only the features presented in this paper. The interesting point is that the nature of our needed glue code is different with respect to the Scala glue code: Scala requires lots of trait multiple inheritance declarations to explicitly merge nested traits with the same name, while in $42_\mu$ we mostly need to add the negative cases for the double dispatch (such as `Sum={method Bool equalToNum(Num that){return false;}}`).

## 6  Summary of formalisation

In Appendix A we formalise $42_\mu$; in addition to conventional soundness, we discuss detailed behaviour and soundness of the compilation process itself; a similar property was called meta-level-soundness in [37]. This property ensures that flattening strictly reduces the number of type errors. In turn, this ensure that reusing a trait cannot induce new type errors. This property was already proved in [37]; here the proof is smoother thanks to our simpler formalisation. Our process requires traits to be well-typed before being reused, however code literals are not required to be well typed before flattening. This design supports mutually recursive types without having to predict the structure of the flattened code, as was needed in [16].

## 7  Related Work

Literature on code reuse is too vast to let us do justice to it in a few pages. In particular, we were unable to discuss all the variations on family polymorphism. Our work is inspired by traits [18], which in turn are inspired by module composition languages [1].

### 7.1  Separating Inheritance and Subtyping

In languages like Cecil [14] and PolyToil [12], classes are not types: it is a more radical solution to 'inheritance implies subtyping', and equivalent to a restricted version of $42_\mu$ where only interface names can be used as types. This complicates typing of **this**, and may prevent any useful application of the **This** type (PolyToil uses polymorphism to support it). Those approaches would ban the following code, since A is not a type:

```
1  class A{ int ma(){return Utils.m(this);} } class Utils{static int m(A a){..}}
```

Cecil syntactic sugar counters this issue.

$42_\mu$ is directly inspired by the 3 independently designed research languages as already mentioned: TraitRecordJ (TR)[6], Package Templates (PT)[26] and Deep-FJig(DJ)[16]. We synthesize the best ideas of those very different designs, while at the same time coming up with a simpler and improved design for separating subclassing from subtyping, which also addresses various limitations of those 3 particular language designs. TR, DJ, and PT are research projects, aiming to be platforms to experiment concepts, not to expose a compact syntax to programmers; instead of using case studies to compare $42_\mu$ against TR, DJ, and PT, in the following we compare various aspects of the language designs on a more theoretical level. We identified 3 properties where one approach shines the most, and 3 properties where one approach is more lacking.

**A simple uniform syntax for code literals.** Between those tree approaches, DJ is best in this sense: TR has separate syntax for class literals, trait literals and record literals. PT is built on top of Java, thus, it must support many different syntactic forms. $42_\mu$ relies on DJ's approach but, thanks to our novel representation of state, $42_\mu$ offers a much simpler and uniform syntax than DJ, TR, and PT.

**Reusable code cannot be "used" (that is be instantiated or used as a type).** This happens in TR and in PT, but not in DJ. In DJ, to allow reusable code to be directly usable, classes introduce nominal types in an unnatural way: the type of `this` is only `This` (sometimes called `<>`) and not the nominal type of its class. That is in DJ 'A=`{method A m(){return this;}}`' is not well typed. This is because 'B= `Use` A' flattens to 'B=`{method A m(){return this;}}`', which is clearly not well typed. Looking at this examples makes it clear why we need reusable code to be agnostic of its name as in TR, PT, and $42_\mu$: either reusable code does not correspond to a type name (as in TR, PT, and $42_\mu$) or all code is reusable and usable, and all code needs to be awkwardly agnostic of its name, as in DJ.

**Requiring abstract signatures is a left over of module composition mindset.** TR and DJ comes from a tradition of functional module composition, where modules are typed in isolation under an environment, and then the composition is performed. As we show in this work, this ends up requiring verbose repetition of abstract signatures which (for highly modularized code) may end up constituting most of the program. Java (and thus PT, as a Java extension) show us a better way: names are understood from their reuse context. The typing of PT offers the same advantages of the $42_\mu$ typing model, but is more indirect. This may be caused by the heavy task of integrating with full Java. Recent work based on TR is trying to address this issue too [17].

**Composition algebra.** The idea of using composition operators over atomic values as in an arithmetic expression is very powerful, and makes it easy to extend languages with more operators. $42_\mu$, DJ, and TR embrace this idea, while PT takes the traditional Java/C++ approach of using an enhanced class/package declaration syntax. The typing strategy of PT also seems to be connected with this decision, so it would be hard to move their approach to a composition algebra setting.

**Complete ontological separation between use and reuse.** While $42_\mu$, TR, DJ, and PT all allow separating inheritance and subtyping only $42_\mu$ and TR properly enforce separation between use (classes and interfaces) and reuse (traits). In DJ all classes are both units of use and reuse (however, subtyping is not induced). PT imports all the complexity of Java: it is possible to separate use and reuse, the model has powerful but non-obvious implications where Java `extends` and PT are used together.

**Naming the self type, even if there is none yet.** TR is lacking here, while $42_\mu$, DJ and PT both allow a class to refer to its name; albeit this is less obvious in PT since both a package and a class have to be introduced to express it. This allows encoding binary methods, expressing patterns like withers or fluent setters and to instantiate instances of (future) classes using the reused code.

## 7.2 Implications for Family Polymorphism

Our **Use** operator is similar to deep mixin composition [19, 22, 41] and family polymorphism [21, 24, 25, 36], but is symmetric while the operator **super** offers flexible explicit conflict resolution. Our presented solution to the expression problem improves over existing solutions in the literature, where one close contender is DJ [16]: our gain over their model is based on our relaxation over abstract signatures. A similar syntax can be achieved with the Scandinavian style [20], or with the work of Nystrom (Jx [31] and J& [32]), where the composition behaves similarly to our sum operator. Both Jx, J&, and the virtual classes of Ernst [21] make use of dependent types. As in .FJ and ^FJ [24, 25, 34, 36], we do not need sophisticated types. The work on DJ [16] contains an in-depth comparison between various FP approaches, including an example written in .FJ syntax synthesizing the difficulty of supporting FP while keeping Use and Reuse connected:

```
1  class A{static class B{int f1;} int k(.B x){ return x.f1;}}
2  class AA extends A{static class B{int f2;} int k(.B x){return x.f2+new .B().f2;}}
```

The syntax .b denotes a relative path, that is, the class b in scope. In FP AA.B further extends A.B: it is implicitly considered a subclass of A.B, adding the field f2. Consider now the following code:

```
1  new AA().k(new AA.B())//well-typed
2  new A().k(new A.B())//well-typed
3  A a=new AA(); //well-typed assuming AA is a subtype of A
4  a.k(new A.B())//runtime error: A.B.f2 does not exist
```

In the sound .FJ type system the last method invocation illtyped even though AA.B is a subtype of A.B. With minor changes, others [11, 21, 25, 32] support this example in the same way. Inheritance implying subtyping is broken only in a controlled way, and it is allowed whenever it does not lead to unsoundness. Recent work on **ThisType** [33, 35] also continues in this line. In those works, "subtyping by subclassing" is preserved: those designs aim to retain the programming model of mainstream OOP languages and backwards compatibility. 42 is instead a radical departure from mainstream OOP, hoping to improve the mechanisms for use and reuse in OOP and unlock new ways to design software.

From a different perspective, we can say that traditional implementations of family polymorphism are still heavily influenced by the "inheritance implies subtyping" model. We believe that this is a major source of complexity in the type systems of those approaches: they need to track calls, and enforce that the *family* of the receiver and the argument is the same. Because we separate inheritance from subtyping we liberate ourselves from tricky issues that arise in such type systems, and can provide a simpler model of family polymorphism, soundly supported by a straightforward nominal type system: by disconnecting use and reuse we outlaw A a=new AA(). In $42_\mu$ this also reduces the expressive power a little, but in the full 42 language, as well as in DJ, the operator *redirect* allows us to write code that is parametric on families of data types. To the same aim, .FJ relies on generics.

Support for FP strictly includes support for the '**This**' type and self instantiation. Scala allows *encoding* further extension/deep mixin composition, but it requires doing it explicitly, growing the amount of required glue-code.

### 7.3 State and traits

The original trait model [18] has no self instantiation and avoids any connection between state and traits. Since it was applied to a dynamic language, the relation with the '**This**' type is unclear.

The idea of abstract state operations emerged from Classless Java [40]. This approach offers a clean solution to handle state in a trait composition setting. Note how abstract state operations are different from just hiding fields under getters and setters: in our model the programmer simply never has to declare what information is stored in fields. The state is computed by the system as an overall result of the whole code composition process.

In the literature there have been many attempts to add state in traits and in module composition languages:

**No initialisation:** The simplest solutions have no constructors, and fields start with **null** (or zero/false). In this setting fields are another kind of (abstract) member, and two fields with identical types can be merged by sum/use; **new** C() can be used for all classes, and init methods may be called later, as in Point p=**new** Point(); p.init(10,30). This approach is commutative and associative. However, objects are created "broken" and the user is trusted with fixing them. While it is easy to add fields, the load of initializing them is on the user; moreover all the objects are intrinsically mutable, preventing a functional programming style.

**Constructors compose fields:** Here a canonical constructor (as in FJ) taking a parameter for each field and just initializing the fields is assumed to be present. It is easy to add fields, however this model (used by [27]) is associative but not commutative: composition order influences field order, and thus the constructor signature. Self instantiation is also not possible since the signature of the constructors change during composition.

**Constructors can be composed if they offer the same exact parameters:** In this model, traits declare fields and constructors initialize their fields using any kind of computation. Traits whose constructors have the same signature can be composed. The composed constructor will execute both constructor bodies in order. This approach is designed in DJ to allow self instantiation. It is associative and mostly commutative: composition order only influences execution order of side effects during construction. However trait composition requires identical constructor signatures: this hampers reuse, and if a field is added, its initial value needs to be synthesized from the other parameters.

### 7.4 Tabular comparison of many approaches

In this table we show if some constructs support certain features: direct instantiation (as in `new C()`), self instantiation (as in `new This()`), is this construct a 'unit of use'?, a 'unit of reuse'?, does using this construct introduce a type? and is the induced type the type of `this`?, support for binary methods, does inheritance of this construct induce subtyping?, is the code of this construct required to be well-typed before being inherited /imported to a new context? is it required to be well-typed before being composed with other code? **Y** and **N** means yes and no; we use "-" where the question is not applicable to the current approach. For example the original trait model was untyped, so typing questions make no sense there.

| | direct instantiation | self instantiation | unit of use | unit of reuse | introduces a type | induced type is the type of this | binary methods | inheritance induces subtyping | typed before imported | typed before composed |
|---|---|---|---|---|---|---|---|---|---|---|
| Java/Scala classes | Y | N | Y | Y | Y | Y | N | Y | Y | N |
| Java8 interfaces | N | N | N | Y | Y | Y | N | Y | Y | N |
| Scala traits | N | N | N | Y | Y | Y | - | Y | Y | N |
| Original traits | N | N | N | Y | - | - | - | N | - | - |
| TR | N | N | N | Y | N | - | N | N | Y | Y |
| $42\mu$ traits | X | Y | X | Y | N | - | Y | N | Y | N |
| $42\mu$ classes | Y | Y | Y | N | Y | Y | Y | - | Y | - |
| Module composition | - | - | Y | Y | - | - | - | - | Y | Y |
| DJ classes | Y | Y | Y | Y | Y | X | Y | N | Y | Y |
| PT | N | Y | N | Y | N | - | - | N | Y | N |

## 8 Conclusions, extensions and practical applications

In this paper we explained a simple model to radically decouple inheritance/code reuse and subtyping. Our decoupling does not make the language more complex: we replace the concept of abstract classes with the concept of traits, while keeping the concepts of interfaces and final classes. Concrete non final classes are simply not needed in our model. Thus, we believe that $42\mu$ is beneficial for code reuse in important cases without having negative impacts on the general programming experience.

The model presented here is easy to extend. More composition operators can be added in addition to `Use`. Variants of the sophisticated operators of DJ are included in the full 42 language. Indeed we can add any operator respecting the following:

- When the operator fails it needs to provide an error that will be reported to the programmer.
- When only well typed code is taken in input, if a result is produced, such result is also well typed.
- When the result is not well typed, the type error must be traced back to a fault in the input.

Our simplified model represents the conceptual core of 42: a novel full blown programming language. In full 42 code literals are first class values, thus we do not need explicit names for traits: they are encoded as methods returning a code literal.

## A  Formalisation

Here we show a simple formalization for the language we presented so far. We also model nested classes, but in order to avoid uninteresting complexities, we assume that all type names are fully qualified from top level, so the examples shown before should be written like: This.Exp, This.Sum, etcetera. In a real language, a simple pre-processor may take care of this step.

In most languages, when implementing an interface, the programmer may avoid repeating abstract methods they do not wish to implement, however to simplify our formalization, we consider source code always containing all the methods imported from interfaces. In a real language, a normalisation process may hide this abstraction[8]. We also consider a binary operator sum (+) instead of the nary operator Use. Figure 1 contains the complete formalization for $42_\mu$: syntax, compilation process, typing, and finally reduction.

### A.1  Syntax

We use $t$ and $C$ to represent trait and class identifiers respectively. A trait ($TD$) or a class ($CD$) declaration can use either a code literal $L$, or a trait expression $E$. Note how in $E$ you can refer to a trait by name. In full 42, we support various operators including the ones presented before and much more, but here we only show the single sum operator: +. This operation is a generalization to the case of nested classes of the simplest and most elegant trait composition operator [18]. Code literals $L$ can be marked as interfaces. We use '?' to represent optional terms. Note that the interface keyword is inside curly brackets, so an uppercase name associated with an interface literal is a interface class, while a lowercase one is a interface trait. Then we have a set of implemented interfaces and a set of member declarations, which can be methods or nested classes. The members of a code literal are a set, thus their order is immaterial. If a code literal implements no interfaces, the concrete syntax omits the implements keyword.

Method declarations $MD$ can be instance methods or static methods. A static method in $42_\mu$ is similar to a static method in Java, but can be abstract. This is very useful in the context of code composition. To denote a method as abstract, instead of an explicit keyword we just omit the implementation $e$.

Finally, expressions $e$ are just variables, instance method calls or static method calls. Having two different kinds of method calls is an artefact of our simplifications. In the full 42 language, type names are a kind of expression whose type helps to model metaclasses. Our values $v_{\overline{D}}$ are just calls to abstract static methods: thanks to abstract state, we have no new expressions, but just factory calls. Thus values are parametric on the shape of the specific programs $\overline{D}$. We then show the evaluation context, the compilation context and full context.

---

[8] In the full 42 language scoping is indeed supported by an initial de-sugaring, and a normalisation phase takes care of importing methods from interfaces.

| | | | |
|---|---|---|---|
| $ID$ | $::=$ | $t \mid C$ | class or trait name |
| $DE$ | $::=$ | $ID = E$ | Meta-declaration |
| $D$ | $::=$ | $ID = L$ | Declaration |
| $E$ | $::=$ | $L \mid t \mid E \ast E \mid \dots$ | Code Expression |
| $L$ | $::=$ | $\{\textbf{interface? implements-}T\ \overline{M}\ \}$ | Code Literal |
| $T$ | $::=$ | $C \mid C.T$ | Type |
| $M$ | $::=$ | $\textbf{static? method } T\ m(\overline{T\ x})\ e? \mid C = L$ | Member |
| $e$ | $::=$ | $x \mid e.m(\overline{e}) \mid T.m(\overline{e})$ | Expression |
| $v_{\overline{D}}$ | $::=$ | $T.m(\overline{v_{\overline{D}}})$, where $m$ is abstract in $\overline{D}(T)$ | value |
| $\mathscr{E}_{\overline{D}}$ | $::=$ | $[\,] \mid \mathscr{E}_{\overline{D}}.m(\overline{e}) \mid v_{\overline{D}}.m(\overline{v_{\overline{D}}}, \mathscr{E}_{\overline{D}}, \overline{e}) \mid T.m(\overline{v_{\overline{D}}}, \mathscr{E}_{\overline{D}}, \overline{e})$ | evaluation context |
| $\mathscr{E}_c$ | $::=$ | $[\,] \mid \mathscr{E}_c.m(\overline{e}) \mid L \ast \mathscr{E}_c \mid \dots$ | compilation context |
| $\mathscr{E}$ | $::=$ | $[\,] \mid \mathscr{E} \ast E \mid E \ast \mathscr{E} \mid \dots$ | ctx |
| $\Gamma$ | $::=$ | $x_1{:}T_1, \dots, x_n{:}T_n$ | variable environment |

(TOP)
$$\frac{E_0 \xrightarrow[\overline{D}]{} E_1 \qquad \forall D \in \overline{D}, \overline{D} \vdash D : \text{OK}}{\overline{D}\ \overline{D'}\ ID{=}E_0\ \overline{DE} \to \overline{D}\ \overline{D'}\ ID{=}E_1\ \overline{DE}}$$

(LOOK-UP)
$$\frac{}{t \xrightarrow[\overline{D}]{} \overline{D}(t)}$$

(CTX-C)
$$\frac{E_0 \xrightarrow[\overline{D}]{} E_1}{\mathscr{E}_c[E_0] \xrightarrow[\overline{D}]{} \mathscr{E}_c[E_1]}$$

(SUM)
$$\frac{}{L_1 \ast L_2 \xrightarrow[\overline{D}]{} L}\ L = L_1 + L_2$$

(CD-OK)
$$\frac{C; \overline{D}, C{=}L_1 \vdash L_1\ :\ \text{OK} \qquad L_1 = L_0[\textbf{This} = C]}{\overline{D} \vdash C{=}L_0\ :\ \text{OK}}\ coherent(C, L_1)$$

(TD-OK)
$$\frac{\textbf{This}; \overline{D}, \textbf{This}{=}L \vdash L\ :\ \text{OK}}{\overline{D} \vdash t{=}L\ :\ \text{OK}}$$

(S-REFL)
$$\frac{}{\overline{D} \vdash T \leq T}$$

(L-OK)
$$\frac{\forall M \in \overline{M}, \quad T; \overline{D} \vdash M : \text{OK}}{T; \overline{D} \vdash \{\_\ \textbf{implements}\ \overline{T}\ \overline{M}\}\ :\ \text{OK}}$$

(NESTED-OK)
$$\frac{T.C; \overline{D} \vdash L\ :\ \text{OK}}{T; \overline{D} \vdash C{=}L\ :\ \text{OK}}$$

(SUBTYPE)
$$\frac{\overline{D} \vdash T_2 \leq T_3 \qquad \overline{D}(T_1) = \{\_\ \textbf{implements}\ \overline{T}\ \_\}}{\overline{D} \vdash T_1 \leq T_3}\ T_2 \in \overline{T}$$

(METHOD-OK)
$$\frac{\text{if } e? = e \text{ then } \overline{D}; \Gamma \vdash e : T_0}{T; \overline{D} \vdash \textbf{static? method } T_0\ m(T_1 x_1 \dots T_n x_n)\ e?\ :\ \text{OK}}$$

if $\textbf{static?} = \textbf{static}$
then $\Gamma = x_1 : T_1\ ..\ x_n : T_n$
else $\Gamma = \textbf{this} : T, x_1 : T_1\ ..\ x_n : T_n$

(SUBSUMPTION)
$$\frac{\overline{D}; \Gamma \vdash e : T_1 \qquad \overline{D} \vdash T_1 \leq T_2}{\overline{D}; \Gamma \vdash e : T_2}$$

(STATIC-METHOD-CALL)
$$\frac{\overline{D}; \Gamma \vdash e_1 : T_1 \ \dots\ \overline{D}; \Gamma \vdash e_n : T_n}{\overline{D}; \Gamma \vdash T_0.m(e_1\ \dots\ e_n) : T}\ \textbf{static method } T\ m(T_1 x_1 \dots T_n x_n)\_ \in \overline{D}(T_0)$$

(X)
$$\frac{}{\overline{D}; \Gamma \vdash x : \Gamma(x)}$$

(METHOD-CALL)
$$\frac{\overline{D}; \Gamma \vdash e_0 : T_0 \ \dots\ \overline{D}; \Gamma \vdash e_n : T_n}{\overline{D}; \Gamma \vdash e_0.m(e_1\ \dots\ e_n) : T}\ \textbf{method } T\ m(T_1 x_1 \dots T_n x_n)\_ \in \overline{D}(T_0)$$

(CTXV)
$$\frac{e_0 \xrightarrow[\overline{D}]{} e_1}{\mathscr{E}_{\overline{D}}[e_0] \xrightarrow[\overline{D}]{} \mathscr{E}_{\overline{D}}[e_1]}$$

(S-M)
$$\frac{}{T.m(\overline{v_{\overline{D}}}) \xrightarrow[\overline{D}]{} meth(\overline{D}(T, m), \overline{v_{\overline{D}}})}$$

(M)
$$\frac{}{v_{\overline{D}}.m(\overline{v_{\overline{D}}}) \xrightarrow[\overline{D}]{} meth(\overline{D}(T, m), v_{\overline{D}}\ \overline{v_{\overline{D}}})}\ v_{\overline{D}} = T.m'(\_)$$

■ **Figure 1** Formalization

## A.2 Well-formedness

The whole program ($\overline{DE}$) is well formed if all the traits and classes at top level have unique names. The special class name **This** is not one of those, and the subtype relations are consistent: this means that the implementation of interfaces is not circular, and that $\forall\ ID\text{=}\mathscr{E}[L] \in \overline{DE}, consistentSubtype(\overline{DE}, \textbf{This}\text{=}L; L)$. That is, every literal declares all the methods declared in its super interfaces. The full 42 language allows covariant return types as in Java. Here for simplicity we require them to have the same type declared in the super interface.

**Define** $consistentSubtype(\overline{DE}; L)$

- $consistentSubtype(\overline{DE}, \{\textbf{interface?implements}\ \overline{T}\ \overline{M}\})$   where
    $\forall T \in \overline{T},\quad \overline{DE}(T) = \{\textbf{interface}\ \_\},$[9]
    $\forall\ \_\text{=}L \in \overline{M},\quad consistentSubtype(\overline{DE}; L)$ and
    $\forall m, T \in \overline{T},\quad$ if $\textbf{method}\ T_0\ m(\overline{T\,x}) \in \overline{DE}(T)$ then $\textbf{method}\ T_0 m(\overline{T\,x})\ e? \in \overline{M}$

A code literal $L$ is well formed iff:

- for all methods: parameters have unique names and no parameter is named **this**,
- all methods in a code literal have unique names,
- all nested classes in a code literal have unique names, and no nested class is called **This**,
- all used variables are in scope, and
- all methods in an interface are abstract, and they contain no static methods.

## A.3 Compilation process

The compilation process is particularly interesting, it includes the flattening process and how and when compilation errors may arise. It is composed by rules TOP, LOOK-UP, CTX-C and SUM. To model more composition operators, they would each need their own rule.

Rule TOP compiles the leftmost top level (trait or class) declaration that needs to be compiled. First it identifies the subset of the program $\overline{D}$ that can already be typed (second premise). Then the expression is executed under the control of such compiled program (first premise). All the traits inside the expression need to be compiled (rule LOOK-UP): $\forall t, \text{if}\ E = \mathscr{E}[t]\ \text{then}\ t \in \text{dom}(\overline{D})$. If the required $\overline{D}$ cannot be typed, this would cause a compilation error at this stage. Rule LOOK-UP replaces a trait name $t$ with the corresponding literal $L$. Since $\overline{D}$ is all well typed, $L$ is well typed too. Rule CTX-C uses the compilation context to apply a deterministic left to right call by value[10] reduction; thus the leftmost invalid sum that is performed will be the one providing the compilation error.

---

[9] That is, in this simplified version in order to implement an interface nested in a different top level name, such interface can not be generated using a trait expression. This limitation is lifted in the full language.

[10] In the flattening process, values are code literals $L$.

Keeping in mind the order of members in a literal is immaterial, rule sum applies the operator:

**Define** $L_1 + L_2$, $\overline{M} + \overline{M}$, $M + M$

- $L_1 + L_2 = L_3$   where
  $L_1 = \{$**interface?** **implements** $\overline{T}_1 \; \overline{M}_1 \; \overline{M}_0\}$
  $L_2 = \{$**interface?** **implements** $\overline{T}_2 \; \overline{M}_2 \; \overline{M}'_0\}$
  $L_3 = \{$**interface?** **implements** $\overline{T}_1, \overline{T}_2 \; \overline{M}_1, \overline{M}_2 \; (\overline{M}_0 + \overline{M}'_0)\}$
  $\mathrm{dom}(\overline{M}_1)$ disjoint $\mathrm{dom}(\overline{M}_2)$ and $\mathrm{dom}(\overline{M}_0) = \mathrm{dom}(\overline{M}'_0)$
- $(M_1 \ldots M_n) + (M'_1 \ldots M'_n) = (M_1 + M'_1) \ldots (M_n + M'_n)$
- $M_1 + M_2 = M_2 + M_1$
- $C$=$L_1 + C$=$L_2 = C$=$L_3$   if $L_1 + L_2 = L_3$
- **static?** **method** $T_0 \; m(\overline{T x}) + $ **static?** **method** $T_0 \; m(\overline{T x})e? = $ **static?** **method** $T_0 \; m(\overline{T x})e?$

Sum composes the content of the arguments by taking the union of their members and the union of their **implements**. Members with the same name are recursively composed. There are three cases where the composition is impossible.

- *Method-clash*: two methods with the same name are composed, but either their headers have different types or they are both implemented.

- *Class-clash*: a class is composed with an interface.[11]

- *Implements-clash*: the resulting code would not be well formed. For example, in the following t1+t2 would result in a class B implementing A with method a(), but B does not have such method.[12]

```
1  t1={ A= {interface method Void a()} }
2  t2={ A= {interface} B= {implements A} }
```

Implements-clash can happen only when composing nested interfaces. Note that while the first two kind of errors are obtained directly by the definition of $L_1 + L_2$, Implements-clash is obtained from well-formedness, since injecting the resulting $L$ in to the program would make it ill-formed by *consistentSubtype*($\overline{DE}, L$).

### A.4 Typing

Typing is composed by rules cd-ok, td-ok, l-ok, nested-ok and method-ok, followed by expression typing rules subsumption, method-call, x and static-method-call.

Rules cd-ok and td-ok are interesting: a top level class is typed by replacing all occurrences of the name '**This**' with the class name $C$, and is required to be coherent. On the other hand, a top level trait is typed by temporarily adding a mapping for **This** to the typed program.

---

[11] The full language relaxes this condition, for example an empty class can be seen as an empty interface during composition.

[12] In 42$\mu$ it could be possible to try to patch class B, for example by adding an abstract method a(); we choose to instead give an error since in the full 42 language such patch would be able to turn coherent private nested classes into abstract (private) ones.

**Define** *coherent*(T, L)

- *coherent*(T, {**interface?** **implements** $\overline{T}$ $\overline{M}$}) holds where
  $\forall C{=}L' \in \overline{M} coherent(T.C, L')$
  and either **interface?** = **interface**
  or $\forall$ **method** $T'$ $m(\overline{T x}) \in \overline{M}$, state(factory(T, $\overline{M}$), **method** $T'$ $m(\overline{T x})$)

A Literal is *coherent* if all the nested classes are coherent, and either the Literal is an interface, there are no static methods, or all the static methods are a valid *state* method of the candidate *factory*. Note, by asking for **method** $T'$ $m(\overline{T x}) \in \overline{M}$ we select only abstract methods.

**Define** factory(T, $\overline{M}$)

- factory(T, $M_1 \ldots M_n$) = $M_i$ = **static method** $T$ $m(\_)$ where
  $\forall j \neq i$. $M_j$ is not of the form **static method** $\_\_(\_)$

The factory is the only static abstract method, and its return type is the nominal type of our class.

**Define** state(M, M′)

- state(**static method** $T$ $m(T_1 x_1 \ldots T_n x_n)$, **method** $T_i$ $x_i()$)
- state(**static method** $T$ $m(T_1 x_1 \ldots T_n x_n)$, **method** $T$ with$x_i(T_i$ that))

A non static method is part of the *abstract state* if it is a valid getter or wither. In this simple formalism without imperative features we do not offer setters.

Rule NESTED-OK helps to accumulate the type of **this** so that rule METHOD-OK can use it. Rule L-OK is so simple since all the checks related to correctly implementing interfaces are delegated to the well formedness criteria. The expression typing rules are straightforward and standard.

## A.5 Formal properties

As can be expected, $42_\mu$ ensures conventional soundness of expression reduction. This property is expressed on a completely flattened program (a program where all $E$ are of form $L$):

**Theorem A.1** (Main Soundness). *if* $\vdash \overline{D} : OK$, *e not of form* $v_{\overline{v}}$ *and* $\overline{D} \vdash e : T$ *then* $e \leftarrow_{\overline{D}} \_$

The proof is standard since the flattened language is just a minor variation over FJ.

In addition to conventional soundness of expression reduction, $42_\mu$ ensures soundness of the compilation process itself. A similar property was called meta-level-soundness in [37]; here we can obtain the same result in a much simpler setting. We denote *wrong*($\overline{D}$, E) to be the number of $Ls$ such that $E = \mathcal{E}[L]$ and not $\overline{D} \vdash L$: OK.

**Theorem A.2** (Compilation Soundness). *if* $E_0 \xrightarrow{\overline{D}} E_1$ *then wrong*($\overline{D}, E_0$) $\geq$ *wrong*($\overline{D}, E_1$).

This can be proved by cases on the applied reduction rule:

- LOOK-UP preserves the number of wrong literals: $t \in \overline{D}$ and $\overline{D}$ is well typed by TOP preconditions.
- SUM either preserves or reduces the number of wrong literals: the core of the proof is to show that the sum of two well typed literals produces a well typed one. A code literal is well typed (L-OK) if all its method bodies are correct. This holds since those same method bodies are well typed in a strictly weaker environment with respect to the one used to type the result. This is because every member in the result of the sum is structurally a subtype of the corresponding members in the operands. Note that by well formedness, if SUM is applied the result still respect *consistentSubtype*.

Compilation Soundness has two important corollaries:

- A class declared without literals is well-typed after flattening; no need of further checking.
- If a class is declared by using literals $L_1 \ldots L_n$, and after successful flattening $C = L$ can not be type-checked, then the issue was originally present in one of $L_1 \ldots L_n$. This also means that as an optimization strategy we may remember what method bodies come from traits and what method bodies come from code literals, and only type-check the latter. If the result can not be type-checked, either it is intrinsically illtyped or a referred type is declared *after* the current class. As we see in the next section, we leverage on this to allow recursive types.

### A.6 Advantages of our compilation process

Our typing discipline is very simple from a formal perspective, and is what distinguishes our approach from a simple minded code composition macros [4] or rigid module composition [1]. It is built on two core ideas:

**1: Traits are *well-typed* before being reused.**   For example:

```
1  t={method int m(){return 2;}
2      method int n(){return this.m()+1;}}
```

t is well typed since m() is declared inside of t, while the following would be illtyped:

```
1  t1={method int n(){return this.m()+1;}} //illtyped
```

**2: Code literals are *not* required to be well-typed before flattening.**   A literal $L$ in a declaration $D$ must be well formed and respect *consistentSubtype*, but it is not type-checked until flattening is complete: only the result is required to be well-typed. For example the following is correct since the result of the flattening is well-typed:

```
1  C= Use t, {method int k(){return this.n()+this.m();}}//correct code
```

The code literal {method int k(){ return this.n()+this.m();}} is not well typed: n, m are not locally defined. This code would fail in many similar works in literature [5, 7, 16] where the literals have to be *self contained*. In this case we would have been forced to declare abstract methods n and m, even if t already provides such methods.

This relaxation allows multiple declarations to be flattened one at the time, without typing them individually, and only typing them all together. In this way, we support recursive types[13] between multiple class declarations without the need of predicting the resulting shape[14].

As seen in TOP, our compilation process proceeds in a top-down fashion, flattening one declaration at a time, a declaration needs to be type-checked where their type is first needed, that is, when they are required to type a trait used in a code expression. That is, in $42_\mu$ typing and flattening are interleaved. We assume our compilation process stops as soon as an error arises. For example:

```
1  ta={method int ma(){return 2;}}
2  tc={method int mc(A a, B b){return b.mb(a);}}
3  A= Use ta
4  B= {method int mb(A a){return a.ma()+1;}}
5  C= Use tc, {method int hello(){return 1;}}
```

In this scenario, since we compile top down, we first need to generate A. To generate A, we need to use ta (but we do not need tc, in rule TOP, $\overline{D}$ = ta and $\overline{D}'$ = tc). At this moment, tc cannot be compiled/checked alone: information about A and B is needed. To modularly ensure well-typedness, we only require ta to be well typed at this stage; if it is not a type-error will be raised immediately. Now, we need to generate C, and hence type-check tc. A is guaranteed to be already type-checked (since it is generated by an expression that does not contain any $L$), and B can be typed. Finally tc can be typed and reused. If the SUM rule could not be performed (for example if tc had a method hello too) a composition error would be generated at this stage. On the other hand, if B and C were swapped, as in:

```
1  C= Use tc, {method int hello(){return 1;}}
2  B= {method int mb(A a){return a.ma()+1;}}
```

we would be unable to type tc, since we need to know the structure of A and B. A type error would be generated.

**The cost: what expressive power we lose**   We require declarations to be provided in the right dependency order, but sometimes no such order exists. An example of a "morally correct" program where no right order exists is the following:

```
1  t= { int mt(A a){return a.ma();}}
2  A= Use t, {int ma(){return 1;}}
```

Here the correctness of t depends on A, that is in turn generated using t. We believe any typing allowing such programs would be fragile with respect to code evolution, and could make human understanding of the code-reuse process much harder. In sharp contrast with others (TR, PT, DJ, but also Java, C#, and Scala) we chose to not support this kind of involved programs.

TR, PT, DJ, Java, C#, and Scala accept a great deal of complexity in order to predict the structural shape of the resulting code before doing the actual code reuse/adapta-tion. Those approaches logically divide the program in groups of mutually dependent classes, where each group may depend on a number of other groups. This forms a direct acyclic graph of groups. To type a group, all depended groups are typed, then the signature/structural shape of all the classes of the group are extracted. Finally, with the information of the dependent groups and the current group, it is possible to type-check the implementation of each class in the group.

---

[13] OO languages leverage on recursive types most of the times: for example String may offer a Int size() method, and Int may offer a String toString() method. This means that typing classes String and Int in isolation one at a time is not possible.

[14] This is needed in full 42: it is impossible to predict the resulting shape since arbitrary code can run at compile time.

In the world of strongly typed languages we are tempted to first check that all will go well, and then perform the flattening. Such methodology would be redundant in our setting: we can only reuse code through trait names; but our point of relaxation is only the code literal: in no way can an error "move around" and be duplicated during the compilation process. That is, our approach allows safe libraries of traits and classes to be typechecked once, and then deployed and reused by multiple clients: as Theorem A.2 states, in $42_\mu$ no type error will emerge from library code.

### A.7 Expression reduction

Our reduction rules are incredibly simple and standard. A great advantage of our compilation model is that expressions are executed on a simple fully flattened program, where all the composition operators have been removed. From the point of view of expression reduction, $42_\mu$ is a simple language of interfaces and final classes, where nested classes give structure to code but have no special semantics. The reduction of expressions is defined by rules CTX-V, S-M, and M. The only interesting point is the auxiliary function $meth$:

**Define** $meth(M, \overline{v_{\overline{b}}})$

- $meth(\textbf{static method } T\ m(T_1\,x_1 \ldots T_n\,x_n)e, v_{\overline{b}1} \ldots v_{\overline{b}n}) = e[x_1 = v_{\overline{b}1}, \ldots, e_n = v_{\overline{b}n}]$
- $meth(\textbf{method } T\ m(T_1\,x_1 \ldots T_n\,x_n)e, v_{\overline{b}0}, \ldots, v_{\overline{b}n}) = e[\textbf{this} = v_{\overline{b}0}, x_1 = v_{\overline{b}1}, \ldots, e_n = v_{\overline{b}n}]$
- $meth(\textbf{method } T_i\ x_i(), T.m(v_{\overline{b}1} \ldots v_{\overline{b}n})) = v_{\overline{b}i}$
  where $\overline{D}(T, m) = \textbf{static method } T\ m(T_1\,x_1 \ldots T_n\,x_n)$
- $meth(\textbf{method } T\ \text{with} x_i(T_i\ \text{that}), T.m(v_{\overline{b}1} \ldots v_{\overline{b}n})\ v_{\overline{b}}) = T.m(v_{\overline{b}1} \ldots v_{\overline{b}i-1}, v_{\overline{b}}, v_{\overline{b}i+1} \ldots v_{\overline{b}n})$
  where $\overline{D}(T, m) = \textbf{static method } T\ m(T_1\,x_1 \ldots T_n\,x_n)$

Here we take care of reading method bodies and preparing for execution. The first case is for static methods and the second is for instance methods. The third and fourth cases are more interesting, since they take care of the abstract state: the third case reduce getters and the fourth reduces withers. In our formalisation we are not modelling state mutation, so there is no case for setters.

We omit the proof of conventional soundness for the reduction. It is unsurprising, since the flattened calculus is like a simplified version of Featherweight Java [23].

**About the authors**

**Hrshikesh Arora** Student at Victoria University of Wellington.
Contact him at arorahrsh@ecs.vuw.ac.nz

**Marco Servetto** Lecturer at Victoria University of Wellington.
Contact him at marco.servetto@ecs.vuw.ac.nz

**Bruno C. d. S. Oliveira** Assistant Professor at The University of
Hong Kong.
Contact him at bruno@cs.hku.hk